# Long Short-Term Memory Neural Networks for False Information Attack Detection in Software-Defined In-Vehicle Network

Zadid Khan, Mashrur Chowdhury, *Senior Member, IEEE*, Mhafuzul Islam, Chin-Ya Huang, and Mizanur Rahman

*Abstract*— **A modern vehicle contains many electronic control units (ECUs), which communicate with each other through the in-vehicle network to ensure vehicle safety and performance. Emerging Connected and Automated Vehicles (CAVs) will have more ECUs and coupling between them due to the vast array of additional sensors, advanced driving features and Vehicle-to-Everything (V2X) connectivity. Due to the connectivity, CAVs will be more vulnerable to remote attackers. In this study, we developed a software-defined in-vehicle Ethernet networking system that provides security against false information attacks. We then created an attack model and attack datasets for false information attacks on brake-related ECUs. After analyzing the attack dataset, we found that the features of the dataset are time-series that have sequential variation patterns. Therefore, we subsequently developed a long short term memory (LSTM) neural network based false information attack/anomaly detection model for the real-time detection of anomalies within the in-vehicle network. This attack detection model can detect false information with an accuracy, precision and recall of 95%, 95% and 87%, respectively, while satisfying the real-time communication and computational requirements.**

## I. INTRODUCTION

TODAY'S vehicles contain many electronic control units (ECUs) that communicate with each other through the in-vehicle network (i.e.- controller area network (CAN), CAN with flexible data-rate (CAN-FD), Flexray) bus to ensure vehicle safety and performance [1]. The modern vehicles have a large number of ECUs due to the addition of new and advanced features such as lane keeping and forward collision warning. Vehicles are also having increased connectivity with other vehicles, infrastructure and the cloud (vehicle-to-everything or V2X), due to the advancement in wireless communication technology. Cellular-V2X is an enhancement of the cellular technology which allows direct device-to-device V2X communication at the physical layer (PC5 for LTE-V) [2]. In addition to Cellular V2X, Bluetooth and Wi-Fi offer low-range alternatives for V2X communication. The vehicle's interface with the outside environment increases the risk of attack upon its own computer systems and networks [3]. For example, connected vehicles (CVs) are vulnerable to compromise from over-the-air software updates containing malware in which attackers gain access to ECUs [4]. As all the ECUs are connected via the in-vehicle network, the attackers can get access to any ECU if they can get access to one ECU. The in-vehicle network is thus open to a variety of vulnerabilities that create an attractive target for attacks, particularly false information injection attacks, which can threaten the safety of a vehicle and those around it [5].

From the perspective of information security, failure to detect false information injection attacks in real-time for safety-critical applications can cause traffic crashes, injuries, and death [6], [7]. For example, during the braking process, if an attacker injects false malicious messages, such as false brake force message into the in-vehicle network, the anti-lock braking system (ABS) will fail, thus resulting in the wheel-lock phenomenon (i.e., an out-of-control vehicle). Similarly, false information received by an in-vehicle electronic parking brakes (EPB) system can wrongly activate parking brakes, and false steering wheel input can trigger the electronic stability control (ESC) unit to activate the vehicle braking system. The legacy in-vehicle networks (LIVN), such as CAN, CAN-FD and Flexray, are also quite vulnerable to attack via eavesdropping [8]. In one such remotely launched CAN bus attack, the attacker eavesdropped upon and corrupted the messages within a single vehicle, thus triggering a recall of 1.4 million other vehicles [4].

The LIVNs are efficient in exchanging messages between ECUs in real-time. However, they have some security vulnerabilities, such as lack of authentication, encryption and absence of source-destination addresses in data frames.

Manuscript received September 21, 2019. This work was supported in part by the Center for Connected Multimodal Mobility (C²M²) (USDOT Tier 1 University Transportation Center.

Z. Khan, M. Chowdhury, M. Islam and M. Rahman are with the Glenn Department of Civil Engineering, Clemson University, Clemson, SC 29634

USA (e-mail: mdzadik@clemson.edu, mac@clemson.edu, mdmhafi@clemson.edu, and mdr@clemson.edu).

C. Y. Huang is with the Department of Electronic and Computer Engineering, National Taiwan University Science and Technology (NTUST), Taipei City, Taiwan 106 (e-mail: chinya@gapps.ntust.edu.tw)

Although cryptographic message authentication on received CAN frames are available to prevent CAN bus message forgery [9], they are characterized by high processing requirements and high message exchange latency, which sacrifices the CAN bus real-time processing and communication performance [10]. Automotive Ethernet is an emerging technology that offers better security features compared to LIVN technologies. It also has several other advantages over LIVNs such as high bandwidth, availability, scalability, being lightweight and cost-effective [11]. Moreover, software-defined networking (SDN) paradigm offers a new technique for better network management through a separation of the control and data planes. SDN offers greater flexibility and resource management to defend against cyberattacks while minimizing network data traffic congestion. Specifically, it simplifies network management tasks in terms of dynamic flow control, has network-wide visibility (global controller), is network programmable and rests upon a simplified data plane [12]. Therefore, in this study, the authors explore an SDN-based in-vehicle network with an Ethernet backbone designed for real-time detection of false information attack on ECUs. SDN helps with the deployment of a data-driven detection model with a global view in addition to global network monitoring. The main reason of using SDN-based in-vehicle Ethernet network instead of a LIVN is that, SDN controller with a global view of the network can better assist the detection module in addition to providing the flexibility in controlling the data flow within the network. We want to deploy the detection module in a system that can control the data flow of the network, so that appropriate steps can be taken against the attacker. Although we do not consider mitigation modules in this study, any mitigation module can be developed and deployed in the SDN application layer and it will have seamless transition with the detection module. Moreover, the SDN application layer allows seamless interaction between modules, which will be useful if we want to integrate another module with the security module. For example, network management module may need to interact with the security module, which would not be possible without the SDN-based in-vehicle Ethernet network. The SDN based in-vehicle ethernet network and the attack detection model can also overlay the existing in-vehicle network, making it feasible to be deployed on top of existing systems. However, it will reduce the effectiveness of the SDN-based in-vehicle Ethernet network by removing its control of the flows between ECUs. Hence, the SDN-based in-vehicle Ethernet network cannot take appropriate mitigation action after detection.

The two major contributions of this study are as follows: (i) We first develop an attack model and create attack datasets for false information attacks on brake-related ECUs in an SDN based in-vehicle Ethernet network; (ii) we develop a false information attack/anomaly detection model using long short-term memory (LSTM) for the real-time detection of anomalies within the in-vehicle networks. The in-vehicle network data frames contain a high number of features where various types of correlation exits within the features. Moreover, each feature has a continuous series of data that changes with time in specific patterns, based on the sensor readings and vehicle conditions.

When a false information attack will happen, the time-series characteristics or the sequential variation patterns will be altered. Therefore, in order to develop an effective detection model, we need a complex multi-variate time-series classification model that can incorporate the correlations among different features while classifying the incoming data as attack or no-attack. From literature, we have found the LSTM neural networks perform better than baseline models with multivariate correlated time-series [13]. Therefore, we use LSTM neural networks in this study for attack detection. The attack detection model is developed in a generic way so that it can function regardless of the underlying in-vehicle network. For example, in a CAN bus, the attack detection module can be deployed in a separate ECU and the ECU can be connected to the bus, giving it access to all CAN frames. However, as mentioned previously, SDN-based in-vehicle Ethernet network offers several advantages over LIVNs when coupled with the attack detection model.

We use Mininet [14] to create the SDN-based in-vehicle Ethernet network and generate CAN data frames using a dataset of raw CAN frames from the CAN-bus of a vehicle to mimic the in-vehicle network behavior. The LSTM neural network is used for time-series classification of the frames for detecting anomalies. Through our analyses, we evaluate the performance of the attack detection model as well as the overall system.

The outline of this paper is detailed as follows. A discussion of the attack detection model and SDN is described in Section II, the SDN-based in-vehicle Ethernet network is detailed in Section III, the study assumptions are detailed in Section IV, and the data description is included in Section V. In Section VI, the false information attack modeling and misbehavior scenarios are described. In Section VII, the LSTM neural network for false information attack detection is described. A system-level evaluation of the SDN-based framework is presented in Section VIII, followed by concluding remarks in Section IX.

## II. RELATED WORK

Much research has been undertaken to mitigate compromised ECU functionality that can occur from inputs from sensors and other ECUs [15]. The consequences of such acts are quite severe given that these compromised systems can cause ECUs to perform dangerous operations (i.e., disabling brakes). In the subsections below, the current state of research on the detection and mitigation of attacks on in-vehicle network and SDN-based network is described.

### A. Attack Detection

Various countermeasures have been proposed to thwart possible ECU attacks. For example, Ao et al. [16] constructed a sensor model to generate the interval of a given variable based on the sensor measurement and the error bound. If the sensor readings are not received within the interval, the sensor is considered attacked. Park et al. [17] applied a system with multiple sensors to measure the same physical variable and used pairwise inconsistencies between sensor readings to detect malicious sensor attacks. Ivanov et al. [18] applied a fusion polyhedron algorithm on multiple sensor measurements to estimate and compare the sensor value with sensor measurements to detect the attack on a sensor. Although they fused multiple sensor readings to determine the sensor value, they ignored the effects of the sensor measurement noise on sensor readings, which resulted in a low-level attack detection accuracy.

A series of studies were undertaken to detect CAN bus-based attacks without scanning the data field of the CAN frames [19], [20], [21]. Moore et al. [19] designed a data-driven anomaly detection algorithm to identify the signal arrival frequency as the feature of CAN frames for the detection of possible anomalies in the CAN bus. Hoppe et al. [20] proposed an anomaly-based intrusion detection algorithm to analyze the value and periodicity characteristics to detect attacks by noting any changes in those message characteristics. Cho et al. [21] extended a previous intrusion detection algorithm by integrating the clock behaviors of ECU transmitters and detected abnormal changes in the criteria to improve the detection efficiency. ECUs can be hacked if over-the-air software updates are interrupted and malware is injected, which can enable remote access to any ECU. Methods that analyze the data field can detect an attack in this scenario. There are several studies that have also explored the use of the data field for anomaly/intrusion detection. Kang and Kang [22] developed a deep neural network-based intrusion detection system for in-vehicle security which uses the probability-based feature vectors extracted from the in-vehicle network packets. Mo et al. [23] developed a statistical anomaly detection model that scans the data field of the CAN frames to detect abnormalities. However, these studies have several limitations. These studies used false information attack model and associated attack datasets from previous studies, they did not create their own dataset and scenarios. Moreover, these studies do not consider the CAN features as time-series data and do not investigate the sequential variation patterns in the CAN features for attack detection. Finally, these studies consider detection models for CAN only, which limits their capabilities for mitigation after detection is done.

### B. Software-Defined Networking (SDN)

SDN is an emerging technology that is now deemed a viable network security solution [12]. There are four advantages of implementing security solutions using SDN, which are summarized in Table I. Although SDN has been extensively used to implement network functions, only a few studies have been conducted to elucidate the security aspects of SDN, particularly security involving the use of SDN [24], [25], [26], [27]. Halba and Mahmoudi [28] presented an SDN based in-vehicle Ethernet network to enable safety and interoperability of data exchange between ECUs, such as, ABS and ESC). In their study, they show that the proposed SDN-based in-vehicle Ethernet network has a fast link failure recovery while maintaining real-time capabilities. The key feature of the SDN-based Ethernet network is that it is independent of ECU technologies and protocols which ensures network interoperability between different ECUs. Thus, security solutions can be developed separately and deployed on top of the SDN controller for any kind of ECU technology.

TABLE I
OVERALL SUMMARY OF SDN FEATURES AND THEIR POTENTIAL CONTRIBUTIONS TO IN-VEHICLE NETWORK SECURITY

| SDN feature | Feature description | Security benefit |
|---|---|---|
| Dynamic flow control | Control (reroute, forward, drop) network flows dynamically | Identify malicious or suspicious network flows and separate them from benign flows |
| Network-wide visibility with centralized control | Monitor network status and flow information globally | Detect network flooding and network anomalies |
| Network programmability | Program network functions | Develop advanced network security applications efficiently and effectively |
| Simplified data plane | Separate control and data plane for simplification | Modify data plane easily to implement security functions |

### III. SDN-BASED IN-VEHICLE ETHERNET NETWORK

In this study, we present a SDN-based in-vehicle Ethernet network and investigate the security aspect of this network. Unlike previous studies emphasizing Denial of Service (DoS), fuzzy and impersonation attacks on the in-vehicle network, here we investigate false information attack injection from a malicious ECU connected to the network.

The SDN based in-vehicle network for false information detection is shown in Figure 1. The white arrows represent the propagation of attack-free data frames, and the red arrows represent the propagation of false data frames. The OpenFlow switch is configured as a forwarding device to forward all data frames to other ECUs regardless of the information within the data frames. It also communicates with the SDN controller through the OpenFlow protocol, allowing the SDN controller to perform network-wide monitoring. As such, the SDN controller acts as the operating system of the data layer. The controller is connected with the application layer through an application programming interface (API), and the application layer contains all the additional capabilities required for the in-vehicle network functionality. The link between the OpenFlow switch and the ECUs is Ethernet-based. For safety-critical and fault-tolerant communication, the Ethernet connection can be a time-triggered Ethernet (TTE) connection [29].

In this paper, we consider two applications in the SDN application layer: simple-switch program, and false information attack detection model. The simple-switch program configures the OpenFlow switch to act as a switch that only forwards the incoming data frames to all other ECUs connected to it. The false information attack detection model investigates LIVN data frames flowing through the OpenFlow switch to identify if there are some anomalies within the data. The detection model training is performed offline because the training phase is time-consuming. The trained model is then deployed for online real-time prediction.

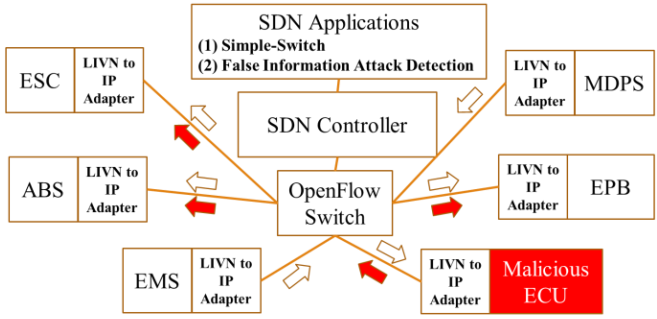

**EMS** = Engine Management System; **ABS** = Anti-lock Braking System; **EPB** = Electronic Parking Brake System; **ESC** = Electronic Stability Control System; **MDPS** = Motor Driven Power Steering; **LIVN** = Legacy In Vehicle Network.

Fig. 1. SDN based in-vehicle Ethernet network system.

In order to show the communication between ECUs effectively, five ECUs are chosen. The number of ECUs is chosen to represent a sub-section of the entire system. The sixth ECU is actually not part of the in-vehicle system but rather a malicious ECU that has connected to the in-vehicle network. ESC, ABS and EPB are brake-related ECUs. We are interested in investigating the data frames which are received by these three ECUs (i.e. ESC, ABS, EPB). Therefore, we have considered two other ECUs, engine management system (EMS) and motor-driven power steering (MDPS) system, which send data to ESC, ABS, and EPB. The actual vehicle network will be an extension of this framework.

Each ECU connects to a universal adapter, which can be termed as a LIVN to IP adapter. Here, CAN is an example of LIVN. The function of the universal adapter is to re-pack the LIVN frames into Ethernet frames and vice-versa. The LIVN to IP adapter extracts the LIVN frames from the transmitted/received data, which contain Ethernet headers, Internet Protocol (IP) headers, Transmission Control Protocol (TCP) headers, LIVN headers, and associated metadata [28]. One OpenFlow switch is connected to five ECUs. These OpenFlow switches receive and send data frames from ECUs using the LIVN to IP adapter. This network can overlay the LIVN and they can co-exist in the same in-vehicle system, because they are two isolated systems with no interconnection between them. However, if the ECUs are connected through LIVN, the SDN controller cannot control the flow of LIVN frames to the ECUs, only attack detection can be achieved. The integration of SDN-based in-vehicle Ethernet network and LIVN frames is an important aspect of this research, as it preserves the functionality of the LIVN while improving the security of the in-vehicle system.

## IV. ASSUMPTIONS OF THIS STUDY

In this section, we detail our five assumptions of the study prior to our description of the analysis, each of which is described below.

**Assumption 1:** A subset of the in-vehicle network is considered to explain the proposed framework effectively.

**Assumption 2:** The training for the LSTM neural network for attack detection occurs offline. The SDN application layer contains the application, which directly predicts the attack status based on data from incoming frames.

| ID | DLC | DATA | Timestamp |
|---|---|---|---|
| 05f0 | 2 | 00 00 0e 00 00 00 00 00 | 2.084334 |
| 04f0 | 8 | 00 00 ff 00 00 00 00 00 | 2.116588 |
| 0690 | 8 | 00 00 00 00 00 00 00 00 | 2.124836 |
| 04f0 | 8 | 00 00 ff 00 00 00 00 00 | 2.126036 |
| 05f0 | 2 | 00 00 0e 00 00 00 00 00 | 2.134353 |
| 0690 | 8 | 00 00 00 00 00 00 00 00 | 2.135407 |
| 04f0 | 8 | 00 00 80 00 eb b6 13 | 2.144996 |
| 0130 | 8 | 00 00 40 ff 00 00 41 3d | 2.156946 |
| 0131 | 8 | 00 00 40 00 00 00 41 9b | 2.157975 |
| 0140 | 8 | 00 00 00 02 29 21 f0 | 2.158382 |
| 04f0 | 8 | 00 00 80 00 eb b6 13 | 2.165245 |
| 0130 | 8 | 00 00 40 ff 00 00 42 1a | 2.176891 |
| 0131 | 8 | 00 00 40 00 00 00 42 bc | 2.177941 |
| 0140 | 8 | 00 00 00 04 25 22 a5 | 2.178345 |
| 05f0 | 2 | 00 00 00 04 25 22 a5 | 2.184367 |
| 04f0 | 8 | 00 00 80 00 eb b6 13 | 2.185408 |
| 0130 | 8 | 00 00 40 00 00 00 43 07 | 2.196924 |
| 0131 | 8 | 00 00 40 00 00 00 43 a1 | 2.197951 |
| 0140 | 8 | 00 00 00 06 26 23 6f | 2.198356 |
| 04f0 | 8 | 00 00 80 00 eb b6 13 | 2.205049 |
| 0430 | 8 | 00 00 00 00 00 00 00 00 | 2.206446 |
| 04b1 | 8 | 00 00 00 00 00 00 00 00 | 2.207510 |
| 01f1 | 8 | 0f 00 00 00 00 00 00 00 | 2.207884 |
| 0153 | 8 | 00 00 ff 00 ff 00 00 00 | 2.208233 |
| 0002 | 8 | 00 00 00 00 00 00 00 50 | 2.208581 |
| 02b0 | 5 | 57 00 00 07 50 08 00 50 | 2.210872 |
| 0002 | 8 | 00 00 01 00 00 08 01 27 | 2.216572 |
| 0153 | 8 | 00 00 ff 00 ff 00 00 00 | 2.217625 |

Fig. 2. Sample CAN bus raw dataset.

**Assumption 3:** The ECUs receive data from different sensors within a vehicle, with the assumption that such data may contain inaccuracies or noise. However, since we found no data related to the sensor noise, these inaccuracies are not considered. The analog-to-digital conversion (converting from raw bytes to actual signal values and vice versa) step may introduce some data errors.

**Assumption 4:** A threshold of 0.7 is used to define the correlation among the features. If two features have an absolute Pearson's correlation coefficient [30] greater than 0.7, then they are considered correlated. The threshold of 0.7 is selected based on a qualitative analysis of the dataset.

**Assumption 5:** Each CAN frame has multiple signal values (values from multiple sensor readings) embedded within the data field. When the attack dataset is created from the attack-free dataset, only one signal value is manipulated at a specific timestamp, with the other signal values of the data remaining unchanged. This means that only the specific bits or bytes of the data are changed.

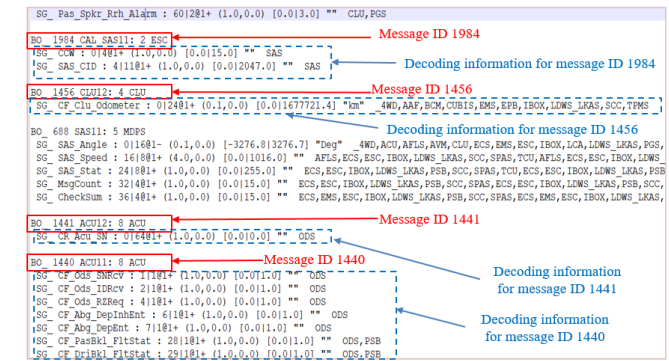

Fig. 3. Generic DBC file for KIA Soul showing the decoded information for message ID 1984, 1456, 1441 and 1440.

TABLE II
PROPERTIES OF FEATURES USED IN THE ATTACK DETECTION MODEL

| Feature no. | Message type | Code (in decimal) | Signal | Bits | Value range | Scale | Offset | Correlations | Maximum deviation |
|---|---|---|---|---|---|---|---|---|---|
| 1 | EMS11 | 790 | TQI_COR_STAT | 4-5 | 0.00-3.00 | 1.00 | 0.00 | 2,4,10,11,14,15 | 0.38 |
| 2 | EMS11 | 790 | TQI_ACOR | 8-15 | 0.00-99.61 | 0.39 | 0.00 | 1,4,10,11,13,14,15 | 0.13 |
| 3 | EMS11 | 790 | N | 16-31 | 0.00-16383.75 | 0.25 | 0.00 | 5,6,10,11,16 | 0.06 |
| 4 | EMS11 | 790 | TQI | 32-39 | 0.00-99.61 | 0.39 | 0.00 | 1,2,10,11,13,14,15 | 0.13 |
| 5 | EMS11 | 790 | TQFR | 40-47 | 0.00-99.61 | 0.39 | 0.00 | 3,6,16 | 0.06 |
| 6 | EMS11 | 790 | VS | 48-55 | 0.00-254.00 | 1.00 | 0.00 | 3,5,16 | 0.30 |
| 7 | EMS12 | 809 | MUL_CODE | 6-7 | 0.00-3.00 | 1.00 | 0.00 | None | 0.54 |
| 8 | EMS12 | 809 | TEMP_ENG | 8-15 | 0.00-143.25 | 0.75 | -48.00 | None | 0.03 |
| 9 | EMS12 | 809 | BRAKE_ACT | 32-33 | 0.00-3.00 | 1.00 | 0.00 | None | 0.37 |
| 10 | EMS12 | 809 | TPS | 40-47 | 0.00-104.69 | 0.47 | -15.02 | 1,2,3,4,11,14,15 | 0.10 |
| 11 | EMS12 | 809 | PV_AV_CAN | 48-55 | 0.00-99.61 | 0.39 | 0.00 | 1,2,3,4,10,14,15 | 0.06 |
| 12 | EMS14 | 1349 | VB | 24-31 | 0.00-25.90 | 0.10 | 0.00 | None | 0.03 |
| 13 | EMS16 | 608 | TQI_MIN | 0-7 | 0.00-99.61 | 0.39 | 0.00 | 2,4,14,15 | 0.11 |
| 14 | EMS16 | 608 | TQI | 8-15 | 0.00-99.61 | 0.39 | 0.00 | 1,2,4,10,11,13,15 | 0.13 |
| 15 | EMS16 | 608 | TQI_TARGET | 16-23 | 0.00-99.61 | 0.39 | 0.00 | 1,2,4,10,11,13,14 | 0.13 |
| 16 | EMS16 | 608 | TQI_MAX | 40-47 | 0.00-99.61 | 0.39 | 0.00 | 3,5,6 | 0.07 |
| 17 | SAS11 | 688 | SAS_ANGLE | 0-15 | 0.00-3276.80 | 0.10 | 0.00 | None | 0.02 |
| 18 | SAS11 | 688 | SAS_SPEED | 16-23 | 0.00-1016.00 | 4.00 | 0.00 | None | 0.74 |
| 19 | SAS11 | 688 | MSGCOUNT | 32-35 | 0.00-15.00 | 1.00 | 0.00 | 20 | 0.45 |
| 20 | SAS11 | 688 | CHECKSUM | 36-39 | 0.00-15.00 | 1.00 | 0.00 | 19 | 0.45 |

## V. DATA DESCRIPTION

The description of the attack free dataset and the correlation analysis of the features are given below.

### A. Attack Free Data

To show the efficacy of our developed system for in-vehicle security, we use the data from real in-vehicle CAN bus. Although other communication protocols, such as FlexRay, are available for the in-vehicle communication network, CAN is a universal and real-time messaging protocol in use in the automotive industry because of its lower implementation costs compared to others [31]. CAN is a broadcast-based network where every ECU can listen to the content of the CAN bus, thus making the bus vulnerable to attacks. After acquiring the CAN frames, the attacker accesses the payload which contains the data field. Previous studies have shown that the acquisition of the CAN data frame makes it possible to create denial of service (DoS) attack, fuzzy attack and impersonation attack on the CAN bus [32]. The CAN bus data used in this study were collected from a previous study on CAN intrusion [32]. The dataset, containing actual CAN data from a KIA soul, was collected by the Hacking and Countermeasure Research Lab [32]. The raw CAN dataset, which contains 537 seconds of data, and 1,048,576 rows of CAN frames, is decoded using a generic DBC file for KIA vehicles. This DBC file is collected from the OpenDBC repository of DBC files created by a startup named Comma.ai. As shown in Figure 2, the dataset contains the CAN ID, DLC (number of bytes in the data), DATA and Timestamp fields.

The DBC file contains the scale and offset values to convert the raw bits of data into signal values. Each CAN frame contains values of different sensor readings, known as signals. The formula used for the conversion is given in (1).

$$Value_{scaled} = Offset + Scale \times Value_{raw} \qquad (1)$$

The dataset is shown in Figure 2, and the DBC file is shown in Figure 3. Note in Figure 3 that each CAN ID has a specific ECU associated with it, which is the origin of the CAN frame. Under each frame, there are multiple signals, each of which has its own specific decoding information (i.e. scale, offset and range). The ECUs which receive and use these signal values are also mentioned here. The EMS and MDPS ECUs send CAN frames to the brake related ECUs: ABS, EPB and ESC. These CAN frames, which contain readings from different in-vehicle sensors, are also received by the malicious ECU. The malicious ECU then manipulates specific bits within the data field and forwards the CAN frames to the brake-related ECUs. As a result, the brake-related ECUs receive both the correct- and false-information signal values. The data field in each CAN frame contains data of several types of signals, which are used as features in the attack detection model. Information on these features is summarized in Table II. Table II contains decoded information from the DBC file, scale, offset, range and correlation information for each signal. The correlation column specifies the other signals which are correlated to it. The details

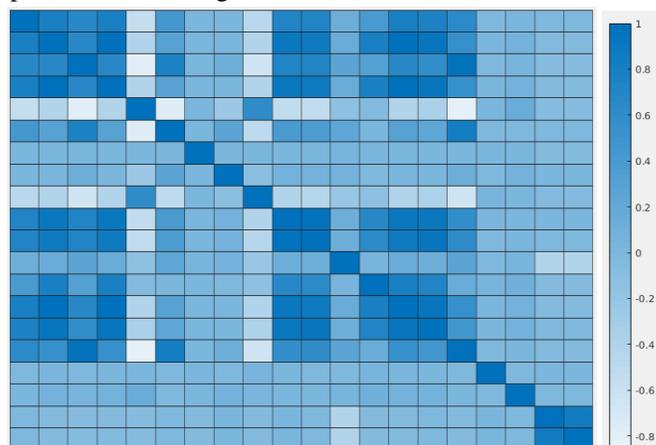

Fig. 4. Correlation heatmap of the feature set.

about the correlation analysis are described in the next subsection.

### B. Correlation Analysis of Attack Free Data

Prior to creating the attack dataset and the detection model, we performed a correlation analysis among the feature set from Table II. The correlation between these signal values is important for the detection of false information. If the value of one feature changes, other correlated feature values should also change.

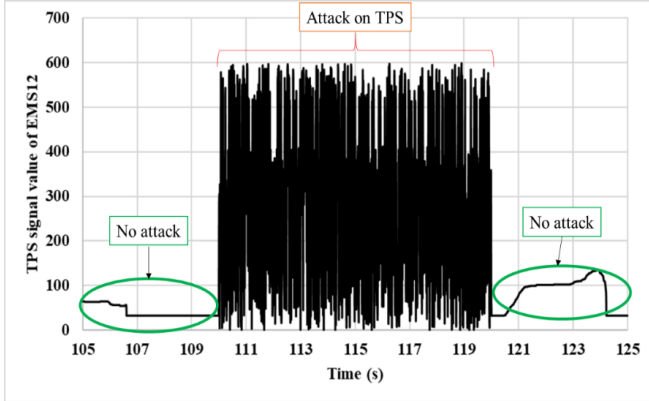

Fig. 5. Attack on TPS Signal of EMS12 ECU.

TABLE III
BYTES ALTERED BY FALSE INFORMATION ATTACKER IN ECUs

| Misbehavior scenarios | Message type | Position of bytes affected | Signal name |
|---|---|---|---|
| 1 | EMS11 | 2 | TQI_ACOR |
| 2 | EMS11 | 3, 4 | N |
| 3 | EMS11 | 6 | TQFR |
| 4 | EMS11 | 7 | VS |
| 5 | EMS12 | 6 | TPS |
| 6 | EMS12 | 7 | PV_AV_CAN |
| 7 | EMS14 | 4 | VB |
| 8 | EMS16 | 1 | TQI_MIN |
| 9 | EMS16 | 2 | TQI |
| 10 | EMS16 | 3 | TQI_TARGET |
| 11 | EMS16 | 6 | TQI_MAX |

However, when there is an attack, other correlated feature values remain unchanged making it possible for the detection model to detect a false information attack. The correlation between features is one of the main reasons for choosing the LSTM neural network. Traditional time-series models analyze time-series as separate entities and do not consider the complex interdependencies among the different time-series. The correlation among the variables has been included in Table II. Let us consider the feature TQFR (Feature no. 5 in Table II), a signal that is correlated with the N, VS and TQI_MAX signals (Feature no. 3, 6 and 16 respectively). The Pearson correlation coefficient is used to determine the correlation, with the color-coded coefficient values shown in Figure 4. The yellow color indicates a strong positive correlation, and the dark blue color indicates a strong negative correlation between the variables. For example, the correlation of TQFR with N, VS and TQI_MAX are -0.80, -0.77, and -0.84 respectively, which indicates a strong negative correlation.

## VI. ATTACK MODELING AND MISBEHAVIOR SCENARIOS

In this section, we describe the false information attack modeling for the in-vehicle CAN networks and the attack scenarios for false information attack that we have used in our study. We first begin with the existing dataset and create an attack dataset based on the false information attacker's model and misbehavior scenarios.

### A. Attack Model

After connecting with the in-vehicle network, an attacker creates a virtual ECU and injects false information within the in-vehicle network. In this context, an attacker acts as a legitimate ECU, receives CAN frames, modifies specific bits within the CAN frames and injects the false CAN frames within the in-vehicle network. This injection of CAN frames can jeopardize the other ECUs, which use these frames. Using the real-world dataset of Kia Soul vehicle, we obtain the dependency of the ECUs and identify the bits within the signal that will be altered to create the attack.

### B. Misbehavior Scenarios

We have created 11 false information attack scenarios based on the dataset as shown in Table III. We created a false information attack with a duration of 10 seconds for each signal. Our dataset contains the Kia Soul CAN bus data for a period of 537 seconds where there exist 1,048,576 CAN data frame from all ECUs. The dataset of 537 seconds is divided into two sets: training dataset and testing dataset. Among all these ECUs data, we consider only the data from the EMS11, EMS12, EMS14, and EMS16 as they are related to the braking and engine control systems. We first analyze the data to select the range of normal behavior from an ECU and generate a random value, which is outside of the three standard deviations of the mean of the error as shown in (2).

$$D_{false} = D_{original} \pm \delta \qquad (2)$$
$$where,$$
$$|\delta| \geq 3\sigma, D_{min} = D_{original} - 3\sigma, and$$
$$D_{max} = D_{original} + 3\sigma$$

where, $D_{false}$ is the false information that is generated by the attacker; $D_{original}$ is the actual value of a signal; $\delta$ is the added value to make $D_{false}$ out of the normal distribution of a signal value; $\sigma$ is the standard deviation or error bound of a signal value; $D_{min}$ and $D_{max}$ are the minimum and maximum value possible value for a signal. To craft the attack data, an attacker creates false information data at ten-second intervals and only changes specific bytes within that 10 seconds. For the testing dataset, the attack time is limited to 5 seconds. Figure 5 shows an example of the false data injection of the 6th byte of EMS12 ECU from the time interval of 110 seconds to 120 seconds. Within this timeframe, the value of the 6th byte of EMS12, which contains the TPS signal information, was randomly selected having $D_{min}=0$, $D_{max}=600$, and $\sigma=0.03$. The attacker then replaces a particular byte of a CAN frame as shown in Table

III.

## VII. LSTM Neural Network for Attack Detection

The data-driven attack detection model resides in the SDN application layer. The model will be pre-trained with the attack dataset. While the system is running, it will detect, in real-time, the occurrence of an attack. The training of the model will occur offline. In this section, we describe the offline aspects of the model, including model development, training and offline testing. As detailed in the previous section, we use 20 signals as input features to the attack detection model. Each input feature is a time-series and the different input features also have some correlation with each other. However, the output of the attack detection model should be a binary value that indicates the current attack status (attack / no attack). Therefore, in this study, we develop an architecture (Figure 6) based on LSTM neural network for time-series classification.

TABLE IV
HYPERPARAMETER TUNING

| Hyperparameter | Values for tuning |
|---|---|
| Number of neurons in LSTM and FC layer | 50, 100 |
| Number of epochs | 50, 200, 500 |
| Batch size | 32, 128, 512, 1024 |
| Dropout probability | 0, 0.5 |

### A. Training and Testing

After the input layer, the first layer of the model is the LSTM layer, which contains multiple LSTM neurons. Compared to the simple recurrent neural network (RNN) neuron, the LSTM neuron has some additional operations that solve the vanishing gradient problem [33]. Its primary feature is the cell state, also known as the memory, which keeps track of the previous timestamps of a time-series. This feature is of particular importance here, as the variations in the input feature have some patterns that are captured by the LSTM neurons. The model can thus detect abnormalities in the input features.

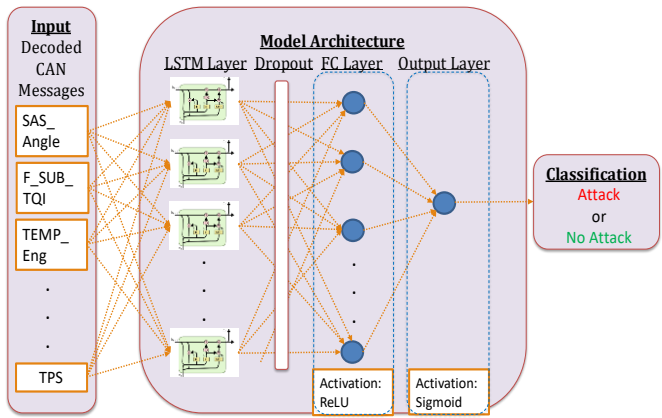

Fig. 6. LSTM neural network architecture for attack detection.

After the LSTM layer, comes the fully connected (FC) layer. The activation function for this layer is the rectified linear unit (ReLU) function, which ensures that there are no negative outputs from any neuron. The link between the LSTM layer and the FC layer contains a dropout probability, which means that the weights of some links will be randomly set to 0. Dropout can reduce overfitting issues of the model.

The final layer is the output layer, which contains only a single neuron. This output of this neuron is binary values of either 0 (i.e., no attack) or 1 (i.e., attack). The activation function for this neuron is the "sigmoid" function, which is the standard activation function for binary classification. The loss function of the model is the binary cross-entropy function and the optimizer used is the Adam optimizer. The formula for binary cross-entropy is given in (3). Here, $H$ denotes the entropy (loss) value, $y_i$ is the binary output and $P(y_i)$ is the predicted probability of that output.

$$H = -\frac{1}{N}\sum_{i=1}^{N}\big(y_i\log\big(P(y_i)\big) + (1-y_i)\log\big(1-P(y_i)\big)\big) \quad (3)$$

The measures of effectiveness considered in this analysis are three standard measures for binary classification: accuracy, precision and recall. These are measures calculated based on the true positive (TP), false positive (FP), true negative (TN) and false negative (FN). Equations (4), (5) and (6) represent accuracy, precision and recall, respectively.

$$Accuracy = \frac{TP+TN}{P+N} \quad (4)$$

$$Precision = \frac{TP}{TP+FP} \quad (5)$$

$$Recall = \frac{TP}{TP+FN} \quad (6)$$

The attack dataset described in the previous section has 537 seconds of data with 1,048,576 CAN frames. At first, we extract a subset of the dataset containing only the CAN frames related to EMS11, EMS12, EMS14, EMS16 and SAS11. This subset of the attack data contains 268,645 rows of data. The dataset is divided into the training set and the testing set. The training set contains data from 0-350 seconds, and the testing set contains the data from 350-537 seconds. In terms of rows, the training set contains 173,914 rows of data, and the testing set contains 94,731 rows of data. The input features in the training and testing set is normalized between 0 and 1 using min-max normalization. As the output is binary, normalization is not required in the output layer.

### B. Hyperparameter Tuning

The model has multiple hyperparameters that require tuning to identify the model with the highest accuracy. In this study, four hyperparameters are used. Their values used for tuning are listed in Table IV. For each hyperparameter combination, we train the model and generate the results in terms of accuracy, precision and recall. The results are summarized using Table V and Table VI. The results in Tables V and VI correspond to the dropout probabilities of 0 and 0.5, respectively.

TABLE VI
ACCURACY, PRECISION AND RECALL FOR DROPOUT=0.5

| Number of neurons | Epochs | Batch size | Accuracy (%) | Precision (%) | Recall (%) |
|---|---|---|---|---|---|
| 50 | 50 | 32 | 93 | 90 | 89 |
| 50 | 50 | 128 | **95** | **95** | 87 |
| 50 | 50 | 512 | 94 | 93 | 88 |
| 50 | 50 | 1024 | 94 | 95 | 87 |
| 50 | 200 | 32 | 93 | 88 | 90 |
| 50 | 200 | 128 | 93 | 87 | 90 |
| 50 | 200 | 512 | 94 | 90 | 90 |
| 50 | 200 | 1024 | 93 | 90 | 89 |
| 50 | 500 | 32 | 93 | 89 | 90 |
| 50 | 500 | 128 | 93 | 88 | 89 |
| 50 | 500 | 512 | 92 | 85 | 90 |
| 50 | 500 | 1024 | 94 | 91 | 89 |
| 100 | 50 | 32 | 93 | 88 | 89 |
| 100 | 50 | 128 | 92 | 86 | 90 |
| 100 | 50 | 512 | 93 | 91 | 88 |
| 100 | 50 | 1024 | 94 | 94 | 87 |
| 100 | 200 | 32 | 93 | 87 | 90 |
| 100 | 200 | 128 | 92 | 85 | 90 |
| 100 | 200 | 512 | 92 | 85 | 90 |
| 100 | 200 | 1024 | 93 | 86 | 91 |
| 100 | 500 | 32 | 92 | 85 | 90 |
| 100 | 500 | 128 | 88 | 76 | **92** |
| 100 | 500 | 512 | 92 | 86 | 90 |
| 100 | 500 | 1024 | 93 | 87 | 90 |

TABLE V
ACCURACY, PRECISION AND RECALL FOR DROPOUT=0

| Number of neurons | Epochs | Batch size | Accuracy (%) | Precision (%) | Recall (%) |
|---|---|---|---|---|---|
| 50 | 50 | 32 | 89 | 78 | 92 |
| 50 | 50 | 128 | 88 | 75 | 93 |
| 50 | 50 | 512 | **94** | **94** | 89 |
| 50 | 50 | 1024 | 93 | 89 | 89 |
| 50 | 200 | 32 | 82 | 65 | 91 |
| 50 | 200 | 128 | 88 | 76 | 92 |
| 50 | 200 | 512 | 83 | 66 | **94** |
| 50 | 200 | 1024 | 89 | 78 | 91 |
| 50 | 500 | 32 | 81 | 63 | **94** |
| 50 | 500 | 128 | 87 | 73 | 92 |
| 50 | 500 | 512 | 81 | 64 | 93 |
| 50 | 500 | 1024 | 83 | 67 | 93 |
| 100 | 50 | 32 | 91 | 83 | 90 |
| 100 | 50 | 128 | 90 | 79 | 93 |
| 100 | 50 | 512 | 91 | 84 | 90 |
| 100 | 50 | 1024 | 92 | 86 | 90 |
| 100 | 200 | 32 | 86 | 71 | 92 |
| 100 | 200 | 128 | 86 | 72 | 93 |
| 100 | 200 | 512 | 85 | 70 | 90 |
| 100 | 200 | 1024 | 86 | 72 | 93 |
| 100 | 500 | 32 | 82 | 65 | **94** |
| 100 | 500 | 128 | 87 | 73 | 93 |
| 100 | 500 | 512 | 87 | 73 | 92 |
| 100 | 500 | 1024 | 88 | 75 | 92 |

Note that the 50-neuron, 50-epoch and 512-batch size model in Table V is the most accurate and the most precise (at 94%). The false-positive rate of 2.45%. Unfortunately, this model has a relatively lower recall score, which means that this model has a higher percentage of false negatives compared to true positives (11%) than the other models. The highest recall achieved with other models is 94%.

Note, as indicated in Table VI, the use of the dropout increases the accuracy of the detection model. The best model includes 50 neurons, 50 epochs and a batch size of 128, with an achieved accuracy and precision score of 95%, a false positive rate of 2.11%. On the other hand, this model has a relatively lower recall score, which means that this model has a higher percentage of false negatives compared to true positives (13%) than the other models. The highest recall achieved with other models is 92%.

## VIII. SYSTEM EVALUATION

We use a Linux virtual machine containing a Ryu controller, an open SDN controller, to create our network topology [34]. We create a Python script for broadcasting and receiving CAN frames, and for measuring the overall communication latency. First, we segment the attack dataset to create a separate dataset for each ECU (hosted in Mininet). We consider five ECUs

(EMS, ABS, ESC, MDPS, and EPB) in addition to a malicious ECU, and five CAN frame types (EMS11, EMS12, EMS14, EMS16 and SAS11) associated with these six ECUs. EMS is solely responsible for broadcasting EMS11, EMS12, EMS14 and EMS16 frames. MDPS is responsible for broadcasting SAS11 frames. It is assumed that the malicious ECU also broadcasts all frames, because it receives all of the data frames forwarded by the OpenFlow switch. The other three ECUs only receive these CAN frames. The refresh rate of all five CAN frame types considered in our attack study is 100 Hz (i.e., 100 frames/seconds). According to the SAE standards, the latency for each periodic message should be less than or equal to the periodicity, in order to maintain real-time operation [35]. Therefore, we establish an overall system latency requirement of less than 10 ms, corresponding to the 100 Hz refresh rate. As indicated in the system-level performance results in Table VII, the average latency at which ABS, EPB and ESC receives the frames is 5.91 ms. The maximum latency observed is 6.05 ms for message type EMS12.

In order to detect the false information in real-time, the detection model must predict within this 10 ms timeframe. We trained and tested the detection model in an Nvidia GeForce GTX 1060 GPU with 3BG memory, after which the model was saved as a JSON file and the model weights were saved as an H5 file. Both the model and its weights were subsequently loaded into the memory and the 50 iterations of the testing times recorded for the same sample point. The average computation time for testing was 1.43 ms, and the average combined latency (communication latency and the computational latency) was 7.34 ms, which is still below the 10 ms threshold. As such, the SDN-based in-vehicle Ethernet network satisfies the real-time requirements. Compared to the CAN bus, this is a major advantage, since the delay in CAN bus varies significantly based on message priority and data traffic on the bus. 92%.

TABLE VII
SYSTEM LEVEL PERFORMANCE OF THE SDN-BASED IN-VEHICLE NETWORK

| ECU | Message type | Data frames | Total time for receiving all frames (s) | Total transmission time (s) | Average latency (ms) | Computational latency (ms) | Total latency (ms) | Latency requirement (ms) |
|---|---|---|---|---|---|---|---|---|
| EMS | EMS11 | 190,555 | 1905.5558 | 1906 | 5.81 | | 7.24 | |
| EMS | EMS12 | 190,554 | 1905.5460 | 1906 | 6.05 | | 7.48 | |
| EMS | EMS14 | 190,554 | 1905.5458 | 1906 | 5.76 | 1.43 | 7.19 | 10 |
| EMS | EMS16 | 190,554 | 1905.5459 | 1906 | 5.93 | | 7.36 | |
| MDPS | SAS11 | 189,884 | 1898.8460 | 1899 | 5.99 | | 7.42 | |
| Overall | - | - | - | - | 5.91 | - | **7.34** | |

## IX. CONCLUSION

In this study, we presented an SDN-based in-vehicle Ethernet network and then investigated its vulnerability to a false information attack. This network can function as the primary in-vehicle network or it can overlay on top of the LIVN. Detection can be achieved using the overlaid architecture, but control of flow will not be achieved. We created an attack dataset and developed a detection model for this specific attack scenario. The attack detection model is generic and can function regardless of the underlying in-vehicle network. Our analyses indicate a 95% success in determining the false information attacks in real-time. We also investigated the overall network/system performance in terms of computation time for detection and overall latency. Results indicate that our attack detection model, developed using the SDN-based in-vehicle Ethernet network, successfully detects false information while meeting the in-vehicle network latency requirement.

While CAN and FlexRay are the most popular technologies for in-vehicle networks, automotive Ethernet technology related to in-vehicle networks can replace them. SDN is well aligned with the automotive Ethernet technology, making it a potential tool for the future in-vehicle network. However, research is required to identify the advantages and limitations of this novel SDN-based in-vehicle Ethernet network concept in emerging in-vehicle networks. This study will provide support in developing new standards for SDN-based in-vehicle Ethernet networks and designing/developing robust attack detection models for these networks. In the future, we will also investigate an attack mitigation module that will receive the attack status information, identify the source of the false information and then request the SDN controller for updating the flow tables of the OpenFlow switch.

## ACKNOWLEDGMENT

This material is based on a study partially supported by the Center for Connected Multimodal Mobility (C$^2$M$^2$) (USDOT Tier 1 University Transportation Center) headquartered at Clemson University, Clemson, South Carolina, USA. Any opinions, findings, and conclusions or recommendations expressed in this material are those of the author(s) and do not necessarily reflect the views of C$^2$M$^2$, and the U.S. Government assumes no liability for the contents or use thereof.

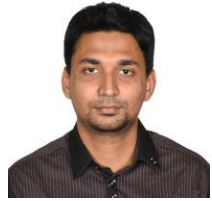

**Zadid Khan** received the B.Sc. degree in electrical and electronic engineering from Bangladesh University of Engineering and Technology (BUET), Dhaka, Bangladesh, in 2014. Then he served as a petroleum engineer in asset development department of Chevron BPC (Bangladesh Profit Center) from 2014 to 2016. After that, he received his M.Sc. degree in civil engineering (transportation major) from Clemson University, Clemson, USA, in 2018. Currently he is pursuing his PhD in the same department. Moreover, he joined the cyber physical systems (CPS) lab as a graduate research assistant from fall 2016, under the supervision of Dr. Mashrur Chowdhury, Professor, dept. of Civil Engineering. His primary research focus is connected and autonomous vehicles (CAVs). Within the CAV domain, his research interests are machine/deep learning, computer networking (SDN, HetNet, Security), data analytics (data fusion, big data), motion planning and parallel / distributed computing.

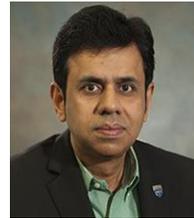

**Mashrur Chowdhury** (SM'12) received the Ph.D. degree in civil engineering from the University of Virginia, USA in 1995. Prior to entering academia in August 2000, he was a Senior ITS Systems Engineer with Iteris Inc. and a Senior Engineer with Bellomo–McGee Inc., where he served as a Consultant to many state and local agencies, and the U.S. Department of Transportation on ITS related projects. He is the Eugene Douglas Mays Professor of Transportation with the Glenn Department of Civil Engineering, Clemson University, SC, USA. He is also a Professor of Automotive Engineering and a Professor of Computer Science at Clemson University. He is the Director of the USDOT Center for Connected Multimodal Mobility (a TIER 1 USDOT University Transportation Center). He is Co-Director of the Complex Systems, Data Analytics and Visualization Institute (CSAVI) at Clemson University. Dr. Chowdhury is the Roadway-Traffic Group lead in the Connected Vehicle Technology Consortium at Clemson University. He is also the Director of the Transportation Cyber-Physical Systems Laboratory at Clemson University. Dr. Chowdhury is a Registered Professional Engineer in Ohio, USA. He serves as an Associate Editor for the IEEE TRANSACTIONS ON INTELLIGENT TRANSPORTATION SYSTEMS. He is a Fellow of the American Society of Civil Engineers and a Senior Member of IEEE.

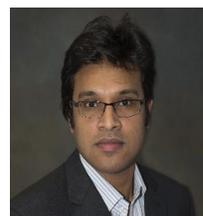

**Mhafuzul Islam** received the BS degree in Computer Science and Engineering from the Bangladesh University of Engineering and Technology in 2014 and MS degree in Civil Engineering from Clemson University in 2018. He is currently a Ph.D. student in the Glenn Department of Civil Engineering at Clemson University. His research interests include Transportation Cyber-Physical

Systems with an emphasis on Data-driven Connected and Autonomous Vehicle.

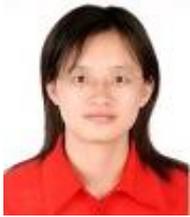

**Chin-Ya Huang** received the B.S. degree in electrical engineering from National Central University, Taiwan, in 2006, the M.S. degree in communication engineering from National Chiao-Tung University, Taiwan, in 2008. She also received M.S. degrees in both electrical & computer engineering (2008) and computer science (2010), and Ph.D degree in electrical & computer engineering (2012) from the University of Wisconsin-Madison. Dr. Huang is currently assistant professor of Electrical and Computer Engineering at the National Taiwan University Science and Technology (NTUST), Taipei, Taiwan. Before joining NTUST in February 2018, Dr. Huang was assistant professor in National Central University from 2013-2018. She was a member of technical staff in Optimum Semiconductor Tech. Inc in New York from 2013-2015. In 2012-2013, she was a senior software engineer at Qualcomm Tech. at San Diego, CA. She also worked as assistant research fellow at Information and Communication Tech. Lab, National Chiao-Tung University, Taiwan in 2013. Her research interest includes 5G, security and machine learning.

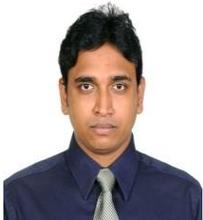

**Mizanur Rahman** received the Ph.D. and M.Sc. degree in civil engineering with transportation systems major in 2018 and 2013, respectively, from Clemson University. Since 2018, he has been a research associate of the Center for Connected Multimodal Mobility (C2M2), a U.S. Department of Transportation Tier 1 University Transportation Center (cecas.clenson.edu/c2m2) at Clemson University. He was closely involved in the development of Clemson University Connected and Autonomous Vehicle Testbed (CU-CAVT). His research focuses on transportation cyber-physical systems for connected and autonomous vehicles and for smart cities.